\documentclass[12pt]{article}
\usepackage{graphicx}
\usepackage{hyperref}
\begin{document}

\newcommand{\be}{\begin{eqnarray}}
\newcommand{\ee}{\end{eqnarray}}
\newcommand{\pl}[1]{Phys. Lett. {\bf #1}}
\newcommand{\zpb}[1]{Zeit. Phys. Lett. {\bf #1}}
\newcommand{\prl}[1]{Phys. Rev. Lett. {\bf #1}}
\newcommand{\pr}[1]{Phys. Rev. {\bf #1}}
\newcommand{\apb}[1]{Ann. of Phys. (N.Y.) {\bf #1} }
\newcommand{\np}[1]{Nucl. Phys. {\bf #1}}
\def\H{{\cal H}}
\def\F{{\cal F}}
\def\L{{\cal L}}
\def\V{{\cal V}}
\def\u{{\bf u}}
\def\k{{\bf k}}
\def\K{{\bf K}}
\def\l{{\bf l}}
\def\Q{{\bf Q}}
\def\d{{\bf d}}
\def\r{{\bf r}}
\def\R{{\bf R}}
\def\v{{\bf v}}
\def\e{{\rm e}}
\def\wb{\varpi}
\def\xb{\bar{\chi}}
\def\K{\rm q}
\def\inti{\int \!\!}

\title{A non-perturbative approach to halo breakup }

\author{J. Margueron $^{a,b}$, A. Bonaccorso $^{a}$, D. M. Brink $^{c}$ \\
 \small $^{a}$ Istituto Nazionale di Fisica Nucleare, Sezione di Pisa, 56127
  Pisa, Italy.\\
\small $^{b}$ GANIL, CEA/DSM-CNRS/IN2P3 BP 5027 F-14076 Caen CEDEX 5, France.\\
\small $^{c}$ Department of Theoretical Physics, 1 Keble Road, Oxford OX1 3NP, U.K.}

\date{\today}

\maketitle

\begin{flushleft} 
 {\bf PACS }
number(s):25.70.Hi, 21.10Gv,25.60Ge,25.70Mn,27.20+n
 \end{flushleft}

\begin{flushleft}  {\bf Key-words}
Nuclear breakup, Coulomb breakup, high order effects, halo nuclei.\end{flushleft}

\begin{abstract}
The theory of weakly bound cluster breakup, like halo nucleus breakup, needs an
accurate treatment of the transitions from bound to continuum states
induced by the nuclear and
Coulomb potentials. When the transition probability  is not very small, a 
non-perturbative framework might be necessary. Nuclear excitation dominates at small
impact parameters whereas the Coulomb potential being long range  
acts over a larger impact parameter interval. In this article, we propose an effective
breakup amplitude which meets a number of  requirements necessary for an
accurate quantitative 
description of the breakup reaction mechanism. Furthermore our treatment gives
some insight on the interplay 
between  time dependent perturbation theory and sudden approximation  and it
allows to include 
the nuclear and Coulomb potentials to all orders   within an eikonal-like
framework. 
\end{abstract}

\section{Introduction}

Break up of halo nuclei is a stimulating field for reaction mechanism studies
where accurate reaction  theories are needed in order to extract spectroscopic information
from experimental data. These theories
should treat together very different excitations induced by electromagnetic and
nuclear fields to all orders. Recently a number of interesting papers
have appeared in which the problem of nuclear and Coulomb breakup
is solved numerically either by a direct  solution of the single particle
Schr\"odinger equation 
\cite{typ01a}-\cite{esb01}, or by DWBA type of approaches \cite{cha02} or by an approximate
treatment of the coupled equations in the continuum \cite{nt}. Of these
papers only Refs.\cite{typ01b,fal01,cha02,nt} have treated at the same time
the nuclear and Coulomb processes. Still one needs analytical models to
complete the understanding of the mechanisms involved and test approximations
which could help reduce the computational difficulties and help extending the
models to more structured clusters. Various analytical solutions of the
Coulomb breakup problem exist where the problem of the higher order effects
has been studied
\cite{typ01a}. In this
paper, we propose an effective amplitude for  breakup to all orders  in the interactions
which reveals the interplay between sudden and time dependent perturbation theory.

Nuclear breakup is a short time process, several observables of  which are
reasonably  well described by the sudden approximation \cite{esb01,bon01}. In the
case of the  electromagnetic field, things are more complicated since the Coulomb
potential is a long range field.  Hence, in  Section 2 of this paper we  present the
theoretical framework for the simultaneous treatment of nuclear and Coulomb
interactions. The effective amplitude we propose is introduced in Section 3 where we
show also how one can establish the accuracy of different approximations based
on the relative behavior of two parameters. The last part of this article
is dedicated to the discussion of some calculations and to their comparison
with experimental data.

\section{Eikonal theory of nuclear and Coulomb breakup}
In a recent paper \cite{mar02} we presented a full description of the treatment of the
scattering equation for a projectile which decays by single neutron breakup due to its
interaction with the target. 
  There it was shown that within the semiclassical approach for the
projectile-target relative motion, the amplitude for a
   transition from a  nucleon bound state $\psi_i$ in the projectile to a final
continuum state
$\psi_f$ is given by 
\begin{equation}
g_{fi}={1\over i\hbar}
\int_{-\infty}^{\infty}dt<\psi_{f} (t)|V({\bf r})|\psi_{i}(t)>,
\label{amp0}
\end{equation}
where $V$ is the interaction responsible for the neutron transition to the continuum. 

For light targets the recoil effect due to the projectile-target Coulomb potential is
rather small  and the interaction responsible for the reaction is mainly the
neutron-target nuclear potential. In the case of heavy targets the dominant reaction
mechanism is Coulomb breakup.  The Coulomb force does not act directly on the neutron but
it affects it only indirectly by causing the recoil of the core.  Therefore the
neutron is subject to an effective force which gives rise to an effective  Coulomb dipole
potential
$V_{\rm  eff}({\bf r,R}(t))$ (cf. Eq.~(\ref{eq4}) of the Appendix). 
${\bf R}(t)={\bf \hat x} d+{\bf \hat z}vt$ is the core-target relative distance and ${\bf
r}$ is the neutron-core coordinate. In ref.\cite{mar02} it was shown that the combined
effect of the nuclear and Coulomb interactions to all orders  can be taken into account by
using the potential
${V}=V_{\rm  nt}+V_{\rm  eff}$ sum of the neutron-target optical potential and the effective Coulomb
dipole potential.   If for the neutron final continuum  wave function we take a distorted
wave of the  eikonal-type, then   the amplitude in the projectile reference
frame  becomes :
\begin{equation}
g_{fi}\left(  \mathbf{k,}\d\right)  = \frac{1}{i\hbar}\inti
d^{3} {\bf r} \int dt \e^{-i{\bf k \cdot r}+i\omega t -i\chi_{\rm  eik}({\bf r},t) }
{V}\left(  {\bf r},t\right)  \phi_{i}\left(  \mathbf{r}\right)
\label{amp1}%
\end{equation}
where $\phi_{i}$ is the time independent part of the neutron initial wave
function and   $i\equiv(l,m)$ stands for the angular momentum quantum numbers,
$\omega=\left(  {{\varepsilon^{\prime}_f}}-\varepsilon_{0}\right)/\hbar
$ and $\varepsilon_0$ is the neutron initial bound state energy  while
${{\varepsilon^{\prime}_f}}$ is the  final neutron-core continuum energy. 
$\k\equiv (k_x,k_y,k_z)$ is a real vector and the eikonal phase shift is
simply
\be
\chi_{\rm  eik}({\bf r},t)=\frac{1}{\hbar v} \int_{t}^\infty V ( {\bf r},t^\prime)
dt^\prime \label{ekph}
\ee
where $v$ is the relative motion velocity at the distance of closest approach. 
Integrating by parts Eq.~(\ref{amp1}) leads to the equivalent expression
for the breakup amplitude :
\be
g_{fi}\left(  \mathbf{k,}\d\right) =-\inti dt \e^{i\omega t} \frac{d}{dt}
\inti d^3{\bf r} \e^{-i \k\cdot{\bf r} -i\chi_{\rm  eik}({\bf r},t)} \phi_{i}\left(
  \mathbf{r}\right).
\label{amp2}
\ee
Eq.~(\ref{amp1}) is appropriate to calculate the coincidence cross section 
$A_p\to (A_p-1)+n$. Finally the
differential probability with respect to the neutron energy and angles can be
written as
$${d^3P_{bu}(d)\over d{{\varepsilon^{\prime}_f}} d\Omega^{\prime}
} ={1\over 8\pi^3}{m k\over \hbar^2}{1\over
2l_i+1}\Sigma_{m_i}| g_{fi}\left(  \mathbf{k,}\d\right)|^2 .\label{anc}$$
where $g_{fi}$ is given by Eq.~(\ref{amp1}) and we have averaged over the neutron
initial state. 

Eq.~(\ref{amp1}) can be in principle an useful alternative to full numerical solutions
of the Schr\"odinger equation. In fact it contains all partial waves in the final
eikonal-like wave function and still the full time dependence, while the numerical
solutions so far available are often restricted to the first few partial waves in the
development of the final continuum wave function. Finally the differential breakup
cross section is given by an integration over core-target impact parameters
\begin{equation}
{d^2\sigma\over  {d{{\varepsilon^{\prime}_f}}d\Omega^{\prime} }}=C^2S
\int_0^{\infty} d\d {d^2 P_{bu}(d)\over
d{{\varepsilon^{\prime}_f}d\Omega^{\prime}}} 
P_{ct}(d), \label{cross} 
\end{equation}
and $C^2S$ is the spectroscopic factor for the initial single particle orbital.  
The effects associated with the core-target interaction have been included by 
multiplying the breakup probability by $P_{ct}(d)=|S_{ct}|^2$ \cite{bon99} the
probability for the core to be left in its ground state, defined in
terms of a core-target S-matrix function of  $d$, the core-target distance
of closest approach. A simple parameterization is
$P_{ct}(d)=\exp(-(\ln 2) \exp[(R_s-d)/a])$, where the strong absorption radius 
$R_s\approx 1.4 (A_p^{1/3}+A_t^{1/3}) $ fm
 is defined as the distance of closest approach
for a trajectory that is 50\% absorbed from the elastic channel and
$a=0.6$ fm is a diffusness parameter.

There have been already in the literature a large number of papers dealing with the
problem of higher order effects in halo breakup and therefore it is important to
understand the relation between our model and other approaches. By
using a first order time dependent amplitude in Eq.~(\ref{amp0}) we are
assuming that  breakup is a one step process in which the neutron is
emitted in the continuum by a single interaction with the nuclear target
potential and by core recoil. The nuclear and Coulomb potential are seen as
final state interactions which distort the simple plane wave which
otherwise would be the final continuum state of the neutron. Because of the
long tail of the halo wave function the overlap between $\psi({\bf r})$ and
the potential
$V({\bf r})$ is large and the potential needs to be treated to all orders.
 This approach is
fully consistent with the usual treatment of higher order effect by the
electromagnetic field \cite{typ01a}.

\section{Approximation scheme : BBM 1 and BBM 2 amplitudes}

Since it is already well established
that the nuclear potential needs to be considered to all orders  for weakly
bound 
projectiles, in
\cite{mar02} we studied numerically only the limits of pure nuclear breakup to
all orders, of 
Coulomb breakup to first order and of the coupling between Coulomb to first
order and nuclear to all orders. 
On the other hand we argued that the question of  if and
when the Coulomb potential needs to be treated to all orders  was still under
investigation.  Here we report  on new calculations that we have recently
performed by  using Eq.~(\ref{amp1}) to treat in detail Coulomb higher order
effects.  

Expanding the time dependent perturbation theory amplitude Eq.~(\ref{amp1}) in
powers of the eikonal
phase shift Eq.(\ref{ekph})
\be
g_{\rm CN}=g_{\rm CN}^{1}+g_{\rm CN}^{2}+g_{\rm CN}^{3}+\dots
\label{eq6}\ee
we get a series of partial amplitudes. From here on we avoid the indices
$_{fi}$ to 
simplify the notation. Treating separately the nuclear and Coulomb potential
Eq.~(\ref{eq6}) 
reduces to 
\be g^{\rm pert}_{\rm C}=g_{\rm C}^{\rm pert\,1}+g_{\rm C}^{\rm pert\,2}+g_{\rm C}^{\rm pert\,3}+\dots \label{eq6abis}\\
g^{\rm pert}_{\rm N}=g_{\rm N}^{\rm pert\,1}+g_{\rm N}^{\rm pert\,2}+g_{\rm N}^{\rm pert\,3}+\dots
\label{eq6bbis}\ee

On the other hand if we make the sudden approximation $\omega=0$  then the analogous amplitudes 
are
\be g^{\rm sudd}_{\rm C}=g_{\rm C}^{\rm sudd\,1}+g_{\rm C}^{\rm
sudd\,2}+g_{\rm C}^{\rm sudd\,3}+\dots \label{eq6atris}\\ g^{\rm
sudd}_{\rm N}=g_{\rm N}^{\rm sudd\,1}+g_{\rm N}^{\rm sudd\,2}+g_{\rm
N}^{\rm sudd\,3}+\dots
\label{eq6btris}\ee
In Eq.~(\ref{eq6atris}) each term corresponds to the  n$^{th}$ term of the standard eikonal
approximation to the theory of Coulomb excitations \cite{ald75}, while in
Eq.~(\ref{eq6abis}) $g_{\rm C}^{\rm pert\,1}$ is the standard first order perturbation
theory. Also we know already that the sudden approximation gives an accurate framework for
the nuclear breakup, then  
$g_{\rm N}^{\rm pert\, i}-g_{\rm N}^{\rm sudd\, i} \approx 0$ for each order, such
that
$g^{\rm sudd}_{\rm N}\approx g^{\rm pert}_{\rm N}$. 

Eq.~(\ref{eq6atris}) is
much easier to calculate than Eq.~(\ref{eq6abis}), however it has the very well known
drawback that the first order term leads to  a logarithmic divergence when
used in the integral over impact parameters Eq.~(\ref{cross}). Under the
hypothesis that higher order terms are accurately calculated by the sudden
approximation, we propose the use of an effective amplitude defined as: 
\be
g_{\rm C}^{\rm BBM\, 1} \equiv  \left( g_{\rm C}^{\rm sudd}
-g_{\rm C}^{\rm sudd\, 1}\right) +g_{\rm C}^{\rm pert\, 1}\label{eq7}
\ee
We want to treat the nuclear process at the same level of approximation,
hence, it is straightforward to show that the generalization of
$g_{ \rm C}^{\rm BBM\, 1}$ which includes also the nuclear potential  to all orders  and the coupling between the nuclear and the Coulomb effective
potential is simply
\be
g_{\rm CN}^{\rm BBM\,1} &\approx& g_{\rm C}^{\rm pert\, 1}+g_{\rm N}^{\rm pert\, 1}
+ \left( g_{\rm CN}^{\rm sudd} - g_{\rm C}^{\rm sudd\, 1}
-g_{\rm N}^{\rm sudd\, 1}\right)\nonumber \\
 &\equiv& g_{\rm C}^{\rm pert\, 1}
+ \left( g_{\rm CN}^{\rm sudd} - g_{\rm C}^{\rm sudd\, 1}
\right).\label{eq8}
\ee
 
In the above equation the choice to treat terms higher than the first order  within
the sudden approximation might seem somewhat arbitrary. Then, we define another
amplitude, BBM 2, for which we keep the time dependence in both first
and second order terms, sudden approximation being used starting from the third
order term~:
\be
g_{\rm  CN}^{\rm BBM\,2} &\equiv& g_{\rm  C}^{\rm pert\, 1}+g_{\rm  C}^{\rm pert\, 2}
+ \left( g_{\rm  CN}^{\rm sudd}-g_{\rm  C}^{\rm sudd\, 1}
-g_{\rm  C}^{\rm sudd\, 2} \right) \nonumber \\
&+&\left( g_{\rm  C}^{\rm \,pert\, 1}-g_{\rm  C}^{\rm sudd\, 1} \right)
g_{\rm  N}^{\rm sudd\, 1}
\ee

\begin{figure}[pht]
\center
\includegraphics[scale=0.5]{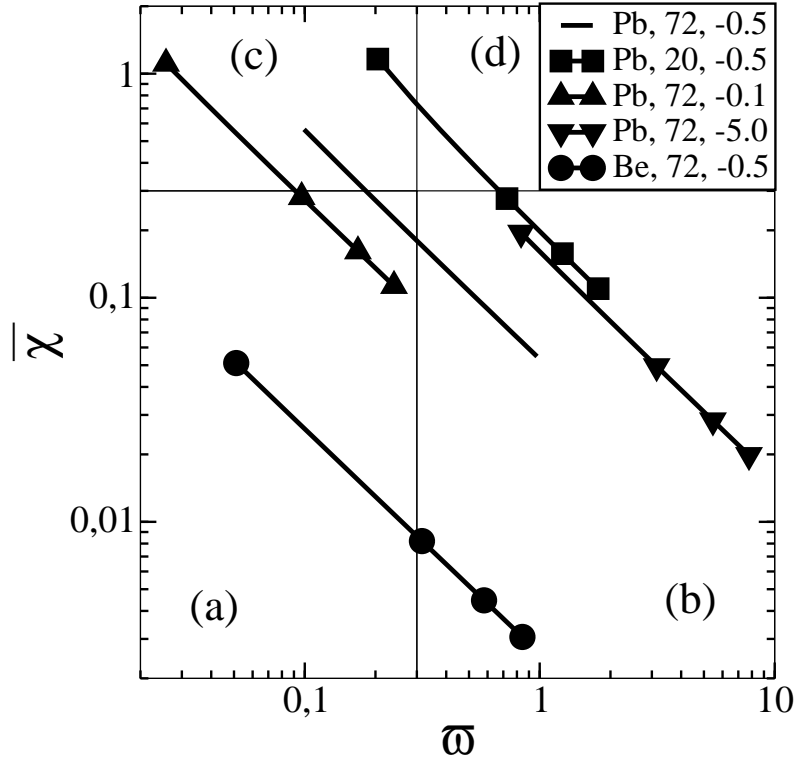}
\caption{Perturbation parameter $\bar \chi$ versus adiabaticity parameter
  $\varpi$ for impact parameters between $R_s$ and 110 fm and for the
  reaction $^{11}$Be+${\rm Target}$. We vary the target (Pb, Be), the energy of
  the beam (72, 20 A.MeV) and the binding energy of the neutron (-0.1, -0.5, -5
  MeV).  Four regions where different theories should be valid are indicated:
  (a) sudden first order, (b) first order time dependent, (c) sudden to all
order, (d) time dependent to
  all orders.}
\label{fig1}
\end{figure}

This approximation scheme solves several problems encountered in the
treatment of halo breakup: the already mentioned logarithmic divergence in the
impact parameter integral due to the the first
order  sudden approximation,
the requirement to treat  the Coulomb and the nuclear field 
at the same level of approximation and to all orders  for small impact
parameters  and the need 
to use time dependent perturbation theory for large impact  parameters where
the Coulomb field 
is effective for a long time. A  quantitative justification of our
approximation scheme is given 
by the discussion of Fig.~(\ref{fig1}) in the next section. On the other hand the
discussion of Fig.~(2) where the results obtained with BBM1 and BBM2 are compared will
clarify the treatment of higher order terms by the eikonal approximation.

\section{Time dependent framework and its sudden approximation}

\smallskip

In this section we give some explicit expressions for the amplitudes discussed in
Sec.(3). We start by considering the Coulomb
term only and in particular the first order approximation for it, thus
$\exp(-i\chi_{\rm   eff})=1$
and $V_{\rm  nt}=0$, 
but the $\omega t$ term is kept in Eq. (\ref{amp2}) (this is the standard first order time
dependent  perturbation theory amplitude) 
\begin{eqnarray}
g^{\rm pert \, 1}_{\rm C}\left( \k,\d;lm \right)    
&=&Q\left( 
\varpi K_{1}\left(  \varpi\right)  \frac{d}{dk_{x} }+i\varpi K_{0}\left(  \varpi\right) 
\frac{d}{dk_{z}}\right)  \tilde{\phi }_{lm}\left(  \mathbf{k}\right) \label{amp8} .
\end{eqnarray}
Here $Q=2{\beta}_{1}Z_{P}Z_{T}e^{2}/\hbar v d$ is the classical Coulomb momentum
transfer to the neutron due to the core recoil. $K_1$ and $K_0$ are the
usual modified Bessel functions. The adiabaticity parameter
$\varpi={{\varepsilon}^{\prime}_f-\varepsilon_{0}\over \hbar v}d$
represents the ratio of the collision time ($d/v$) over the nuclear
interaction time. If the reaction mechanism is such that   $\varpi$ is
small, then the nuclear interaction time is greater than the collision time
and the  sudden approximation becomes accurate.

\smallskip
Then we consider the sudden approximation in which  
$\omega=0$ and Eq.~(\ref{amp1}) can be calculated with  the nuclear and Coulomb
potentials to all orders  giving 
\begin{equation}
g^{\rm sudd}_{\rm CN}\left( \k,\d;lm \right)  =\int d^{2}\mathbf{r}_{\perp}
\e^{-i\mathbf{k}_{\perp}\cdot\mathbf{r}_{\perp}}
\left( \e^{-i(\chi _{\rm nt}\left( \r_\perp \right) 
+\chi_{\rm eff}\left( \r_\perp \right))} -1 \right) 
\tilde \phi_{lm}\left(  \mathbf{r}_{\perp},k_{z}\right),\label{a}
\end{equation}
where
\begin{eqnarray}
\chi_{\rm eff}\left( \r_\perp \right) &=&
\int_{-\infty}^{\infty}dt~~V_{\rm eff}({\bf r,R}(t)) =Qx
\label{C}
\end{eqnarray} 
and $V_{\rm eff}$ is given by Eq.~(\ref{eq4}).
However as we mentioned above, in a first step we study only the effects of
the Coulomb potential. Thus we call $g_{\rm  C}^{\rm sudd}$ 
the  amplitude obtained from Eq.~(\ref{a}), by setting the nuclear
potential  equal to zero. 
Then  the Coulomb amplitude in the sudden approximation to all orders  can be written
as
\begin{eqnarray}
g^{\rm sudd}_{\rm C}\left( \k,\d;lm \right) & =&\int
d^{2}\mathbf{r}_{\perp}e^{-i\mathbf{k}_{\perp}\mathbf{.r}_{\perp}}
\left( e{^{-iQx}}-1\right)
\tilde \phi_{lm}\left(
\mathbf{r}_{\perp},k_{z}\right)\nonumber \\
&=&\tilde{\phi}_{lm}\left( k_{x}+Q,k_{y},k_{z}\right) -\tilde{\phi}
_{lm}\left( k_{x},k_{y},k_{z}\right) \label{d}
\end{eqnarray}

\smallskip
In the limit of very small $Q$  Eq.~(\ref{d}) gives
\begin{equation}
%g^{(1)}_{Csudd}\left( \mathbf{k,}\d\right) \approx Q\frac
g^{\rm sudd \, 1}_{\rm C}\left( \k,\d;lm \right) \approx Q\frac
{d}{dk_{x}}\tilde{\phi}_{lm}\left( k_{x},k_{y},k_{z}\right)\label{b}
\end{equation}
which  is the sudden approximation restricted to first order and it agrees with the
perturbation formula in the sudden limit, because $\varpi K_{1}\left(\varpi\right) =1$ and
$\varpi K_{0}\left(\varpi\right)=0$ when
$\varpi=0$.   
If the initial state wave function is approximated by its asymptotic form which is an
Hankel function 

   \begin{equation}\phi_{lm}(\r)=-i^lC_i\gamma_0h_{l}^{(1)}(i\gamma_0r)
   Y_{lm}(\theta,\phi), \,\,\,\gamma_0 r >>1,
   \label{in}
   \end{equation} 
where $C_i$ is the asymptotic normalization constant
 and $\gamma_0=\sqrt{-2m\varepsilon_0}/\hbar$, 
then the general
form of the initial state momentum distribution is given by the Fourier transform of
Eq.(\ref{in}) :
\begin{equation} \widetilde {\phi}_{lm}({\bf k}) = 4\pi  C_i \frac{(k/\gamma_0)^l}
{(\k^2+\gamma_0^2)}Y_{lm} (\hat{\k}) 
\label{e}
\end{equation} 
For the 2s$_{1/2}$ halo state of $^{11}Be$ it reads
\begin{equation} 
\widetilde {\phi}_{00}(k)=2 \sqrt \pi  C_i {1\over
(\k^2+\gamma_0^2)}=2 \sqrt \pi  C_i {1\over
K^2}\label{f}  
\end{equation}
Where we have put $K^2=\k^2+\gamma_0^2$. By defining the dimensionless strength
parameter $\xb= {Q/K}$  and using Eq.~(\ref{f}), the sudden to all orders amplitude
calculated explicitly up to second order in $\xb$ reads
\begin{eqnarray}
%g_{Csudd}^{all-ord}\left( \mathbf{k,}\d\right)    
g^{\rm sudd}_{\rm C}\left( \k,\d;00\right)    
&= &-{2\sqrt{\pi}C_i\over K^2}
\left[{2k_x\over K}\xb+\left (1-{4k_x^2\over K^2}\right )\xb^2+...\right]\label{g}
\end{eqnarray}

In  Appendix A, Eqs.(\ref{eq40}) and (\ref{eq41}), we give the
expansion of the amplitude up to second order in the full time dependent approach.
Eq.~(\ref{g}), can also be obtained from those equations  in the limit
$\varpi
\to 0$.  The strength parameter $\xb$ represent the ratio of the classical
Coulomb momentum transfer $Q$ over the  momentum $K$ which is an average
of the neutron final and initial momenta. If
$\xb$ is small, then a first order  theory is accurate. From Eq.~(\ref{g})
one sees that if
$k_x=0$, which happens for example for scattering at zero degrees, the amplitude gets
contribution starting from the second order term. Thus the higher order terms
are important for a proper description of the forward angle neutrons.

The transition from the perturbative to the non-perturbative but all orders regime can be
studied by plotting the strength parameter $\xb$ vs
the adiabaticity parameter $\varpi$ using the impact parameter as a variable. We show
in Fig.~(\ref{fig1}) the results obtained for various combinations of incident beam
energy and neutron separation energy and different targets as indicated in the figure.
The case of low incident energy clearly needs an exact treatment of Coulomb breakup
because at all impact parameters both the strength parameter $\xb$ as well as the
adiabaticity parameter $\varpi$ get values close to one. In the other cases instead,
there is always one of the two limits which works well. For small separation energies
(0.1MeV)
$\varpi$ is very small and one can use the sudden approximation to all orders. On
the other hand for large separation energies (5MeV)
$\xb$ is always small and the first order perturbation theory is accurate enough. In the
cases we are discussing in this paper there is  a smooth transition from one regime to the
other and the transition occurs for impact parameters such that $\xb \approx \varpi$
which is satisfied in our case for 
$d=d_{\rm crit}=\sqrt{Z_{\rm eff}/\sqrt{2}\gamma_0|\epsilon_0|}$.
This discussion and our formulation of the sudden approximation are very
close in spirit to the work of Typel and Baur \cite{typ95} %-\cite{typ94a}
and to ref.\cite{tok}.

\begin{table}[t]
\footnotesize
\caption{ For several reactions discussed in this paper, we give the
associated strong absorption radius $R_s$ and the critical impact parameter
$d_{\rm crit}$. The maximum value of the impact parameter 
%assuming that the grazing angle for the core is about 3 deg 
is always $d=120$ fm.} \vskip .1in
%\footnotesize
\begin{center}
\begin{tabular}{ccccc}
\hline\
& Target & $E_{\rm inc}$ (A.MeV) & $R_s$ (fm) & $d_{\rm crit}$ (fm)  \\
\hline
$^{11}$Be & Be & 41 & 6.0 & 4.4   \\
          & Ti & 41 & 8.2 & 10.2 \\
          & Au & 41 & 11.3& 19.4  \\
          & Pb & 72 & 11.4& 19.8  \\
$^{19}$C  & Pb & 67 & 12.0& 18.4  \\
\hline
\end{tabular}
\end{center}
\label{table:1}
\end{table}

\section{Results}

\begin{figure}[pht]
\center
\includegraphics[scale=0.5]{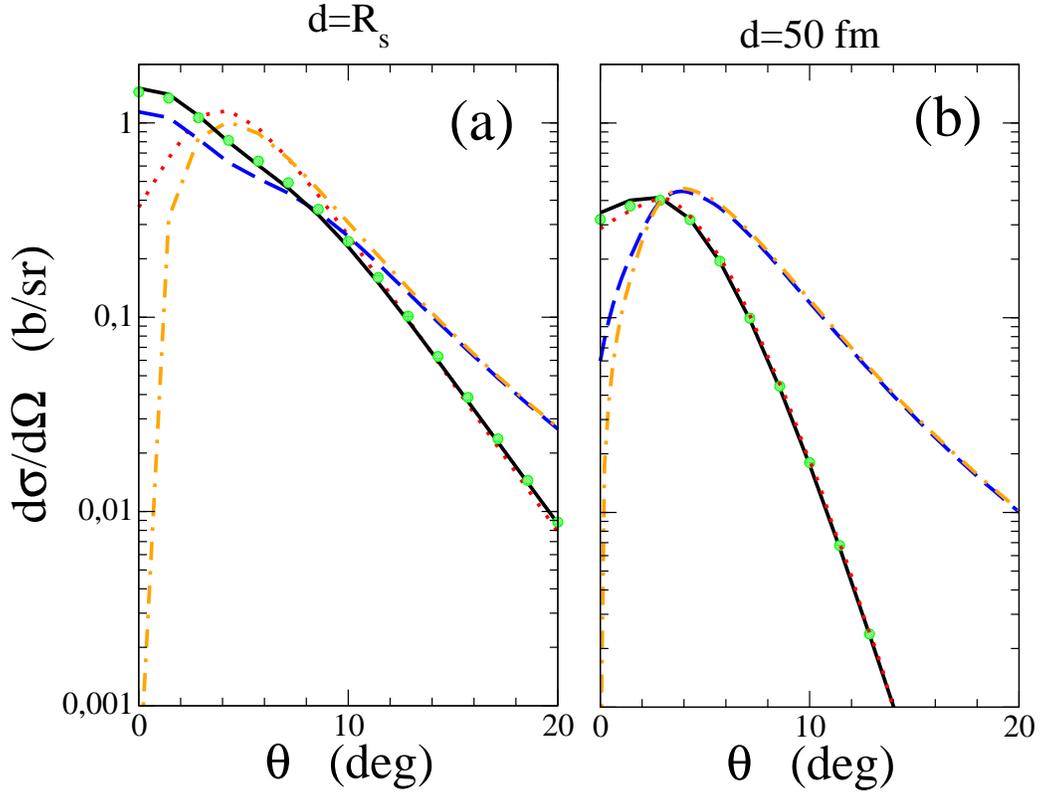}
\caption{Neutron angular distributions for the reaction $^{11}$Be+$^{197}$Au
  at 41 A.MeV at $d=R_s$ and 50 fm.  Solid line is the result of $g_{\rm C}^{\rm BBM\, 1}$
  compared to $g_{\rm CN}^{\rm BBM\, 2}$ calculated without nuclear potential (big dots).
The dotted line is the first order
  perturbation theory calculation, dashed line is the sudden to all orders while dot-dashed line is the sudden to first order result.}
\label{fig2}
\end{figure}

We discuss now the behavior of $g_{\rm C}^{\rm BBM \, 1}$.
An interesting observable which has been measured in a few experiments
\cite{ann94,nak3} is the exclusive neutron angular distribution. The data of
\cite{ann94} were taken at 41 
A.MeV therefore it would be interesting to check if our model works
reasonably well at such low energy. We show in Fig.~(\ref{fig2}) 
(a) and (b) angular distributions for the Coulomb breakup alone calculated for the reaction
$^{11}$Be+$^{197}$Au at 41 A.MeV at two impact parameters $d=R_s$,  and $d=50$
fm. The solid line is the result obtained from $g_{\rm C}^{\rm BBM\, 1}$ which is compared
to the result from
$g_{\rm CN}^{\rm BBM\, 2}$ without nuclear interaction (big dots), the dotted line is the
first order perturbation theory calculation, dashed line is the sudden to all orders while
dot-dashed line is the sudden to first order result. 
We see that at small impact parameters the sudden to first order and the
sudden to all orders  give coincident results, starting from an angle of
about 10 deg. At $d=50$ fm  already 
the sudden to all orders  and the sudden to first order give the same results
at all angles. Thus it 
appears that higher order effects are important only at small neutron angles
and at small impact 
parameters. Also in these situations the $g_{\rm C}^{\rm BBM\, 1}$ amplitude
Eq.~(\ref{eq7}) reduces to the first order perturbation theory amplitude Eq.~(\ref{amp8}).
Eq.~(\ref{eq7}) can be considered correct only if  the
treatment of higher order terms 
by the sudden approximation, is accurate. This is indeed shown
in Fig.~(\ref{fig2}) by the results obtained with the effective formula
corrected to second order $g_{\rm CN}^{\rm BBM \, 2}$ (big dots) calculated without
nuclear potential.

\begin{figure}[pht]
\center
\includegraphics[scale=0.5]{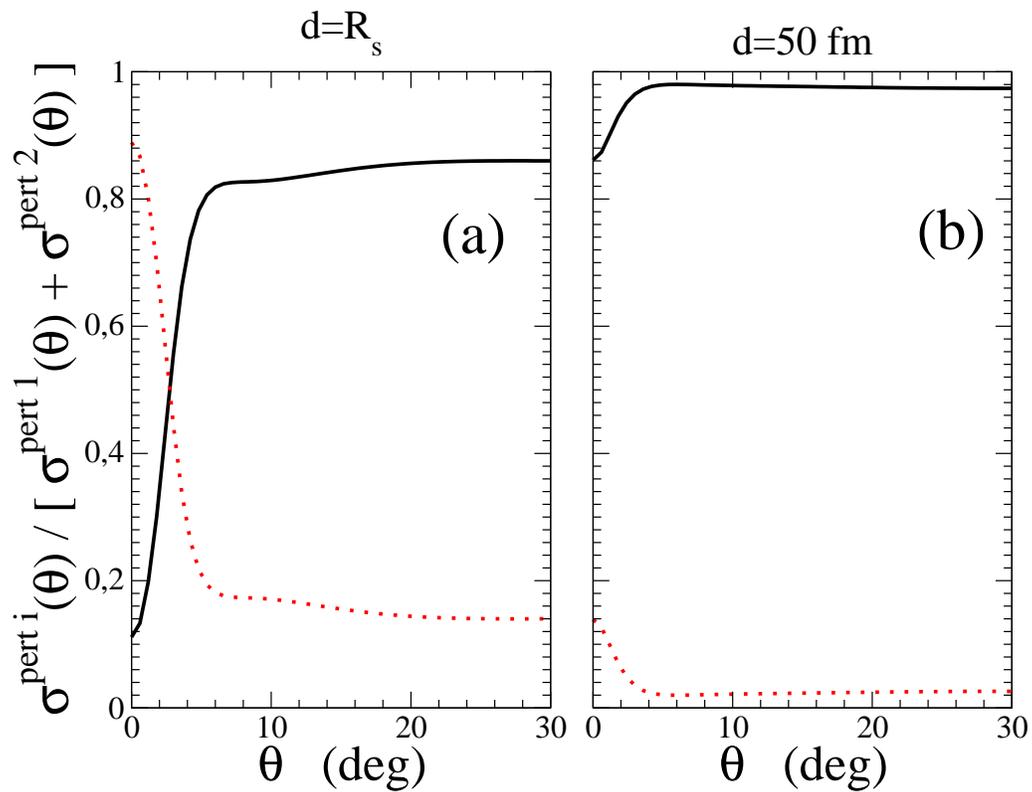}
\caption{Same reaction  as Fig.~(2), the lines are ratios of differential cross
sections as a function of neutron angle. The solid and dotted lines  are  the ratios
of the first and second order time dependent perturbation theory cross
sections respectively  divided by their sum. }
\label{fig4}
\end{figure}

At angles  smaller than  10 deg and small impact parameters the all
 orders calculation is different from the first order
calculation. 
%due to the fact that the sudden first order amplitude, proportional to
%$k_x$ gives zero or very little scattering at small angles.  
%This makes the sudden first order amplitude inadequate to describe
%Coulomb breakup. In this situation our effective formula substitute the first
%order term calculated in the 
%sudden approximation, with the first order calculated in time dependent approach. 
This is due to higher orders  terms in the Coulomb field. Fig.~(\ref{fig4})
clarify this point: we represent the contribution of the first and the
second order calculation normalized by their  sum. It is clear that second order is very
important at small neutron angles for small impact parameters whereas at large impact
parameters, its effects are negligible. 

\begin{figure}[pht]
\center
\includegraphics[scale=0.5]{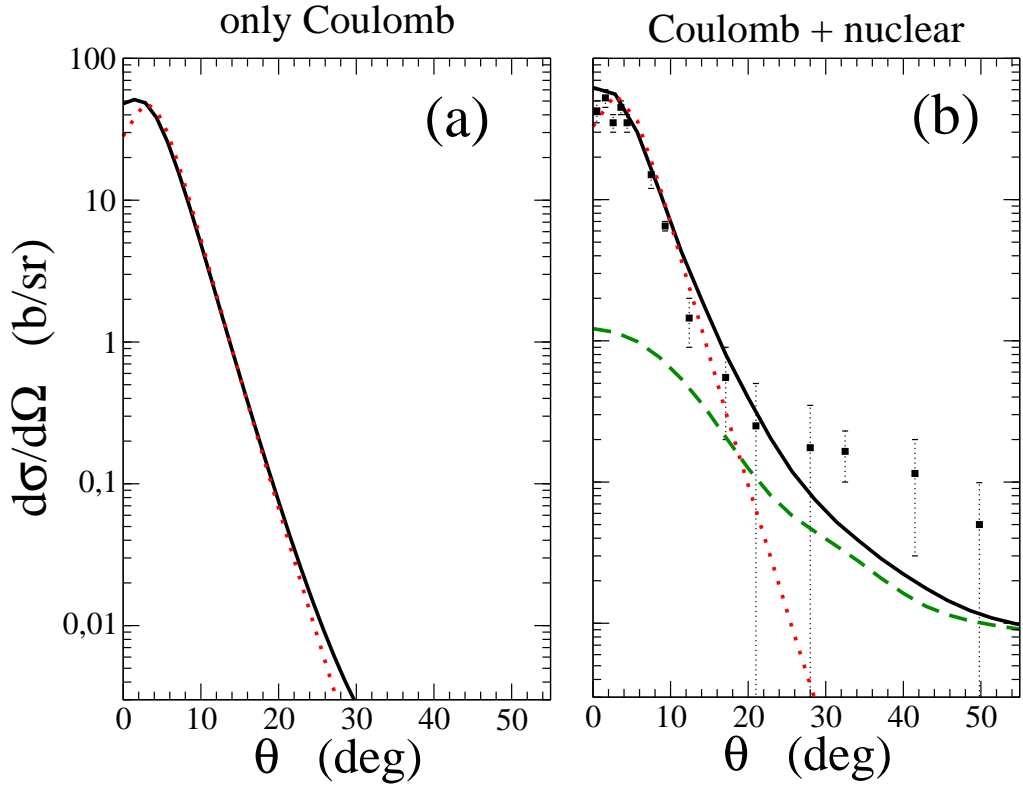}
\caption{Same reaction  as Fig.~(2) with the full  d-integration. 
In both figures the dotted line is first
order perturbation theory. On the left hand
side figure, the solid line is the calculation with the effective amplitude
including only the Coulomb potential, while on the right hand side the solid line
includes both nuclear and Coulomb effects and the dashed line is the
nuclear breakup alone.  Data are from
\cite{ann94}.}
\label{fig3}
\end{figure}

The angular distributions integrated over impact parameters are shown in
Fig.~(\ref{fig3}) (a) and (b). The result with the $g_{\rm C}^{\rm BBM \, 1}$ amplitude in
(a) is shown by the full line while the result with the  Coulomb first order amplitude is
the dotted line. In the case of pure Coulomb breakup, the first order
amplitude is accurate enough down to 10 degrees, where higher order terms in the Coulomb
field flatten the angular distribution.
Such behavior seems indeed to be present in the experimental
data which are shown on the right hand side figure together with the nuclear
contribution. Our results explain why first order perturbation theory has been
so successful in earlier studies of Coulomb breakup and they provide also
a further justification of our previous approach \cite{mar02}. 
Note that for lighter targets the data do show a peak
slightly shifted from zero degrees, thus reflecting the less important
effect of higher order contributions \cite{ann94}.   
In Fig.~(\ref{fig3})  (b) we give instead by the solid line the results from
$g_{\rm CN}^{\rm BBM\, 1}$ including Coulomb and nuclear potential. For completeness we give
also by the dashed line the nuclear contribution alone and the experimental data from
\cite{ann94}.

\begin{figure}[pht]
\center
\includegraphics[scale=0.5]{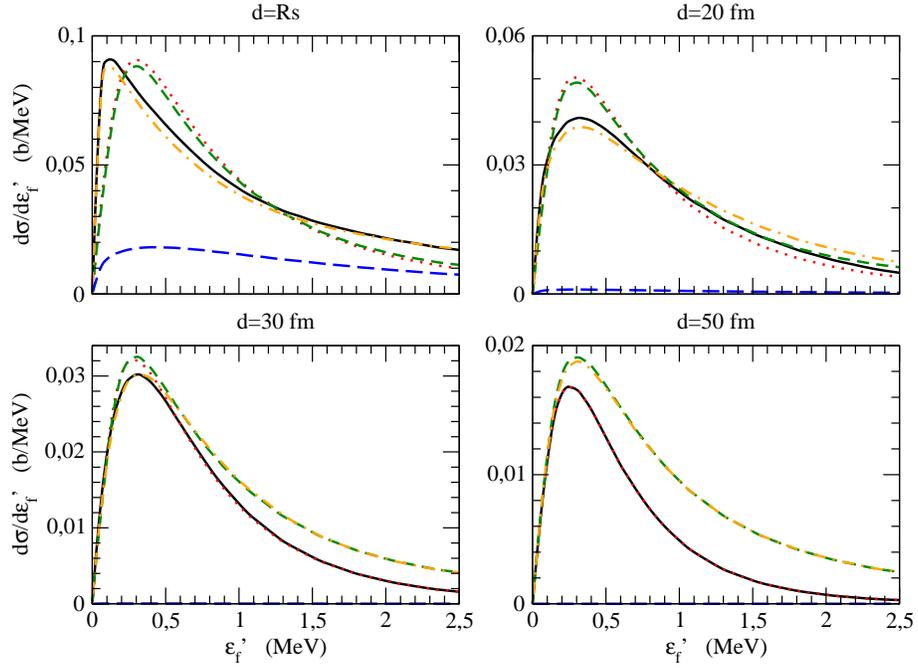}
\caption{Neutron-target energy distributions from $^{11}$Be+$^{208}$Pb at 72 A.MeV for
several impact parameters. Solid line is the result of $g_{\rm CN}^{\rm BBM\, 1}$, the
dotted line is the first order perturbation theory calculation, dashed line is the first
order sudden  calculation, the long dashed is the nuclear sudden to all orders and
dot-dashed line is the Coulomb plus nuclear sudden to all orders result.}
\label{fig5}
\end{figure}

We present now results relative to energy distributions obtained 
for the reaction $^{11}$Be($^{208}$Pb,$^{208}$Pb)$^{10}$Be+$n$ at
$E_{\rm inc}$=72 A.MeV \cite{nak94}. In Fig.~(\ref{fig5}) we give the results of
the calculations obtained by the amplitude $g_{\rm CN}^{\rm BBM\, 1}$, including both
nuclear and Coulomb potential, Eq.~(\ref{eq8}),   by the solid line. The sudden all-order
amplitude (dot-dashed line) Eq.~(\ref{a}), the first order perturbation theory (dotted
curve) Eq.~(\ref{amp8}) and the sudden first order (short dashed) Eq.~(\ref{b}) for an
impact parameter corresponding to the strong absorption  radius $d=R_s$ and  to $d$=20, 30
and 50 fm.  The long dashed line gives only the nuclear contribution.
At small impact parameters the two first order calculations, sudden and time dependent
perturbation theory are very close, thus showing that the sudden approximation is valid at
small impact parameters and therefore  suggesting that Eq.~(\ref{d})  
would be accurate at
small impact parameters to calculate the all orders amplitude.
On the other hand we see that starting from $d=30$ fm a new regime applies in
which the two sudden calculations, to all orders  or to first order give the
same results. Then we conclude again  that at high impact parameters the higher order
effects can be neglected and first order perturbation theory applies.

\begin{figure}[pht]
\center
\includegraphics[scale=0.5]{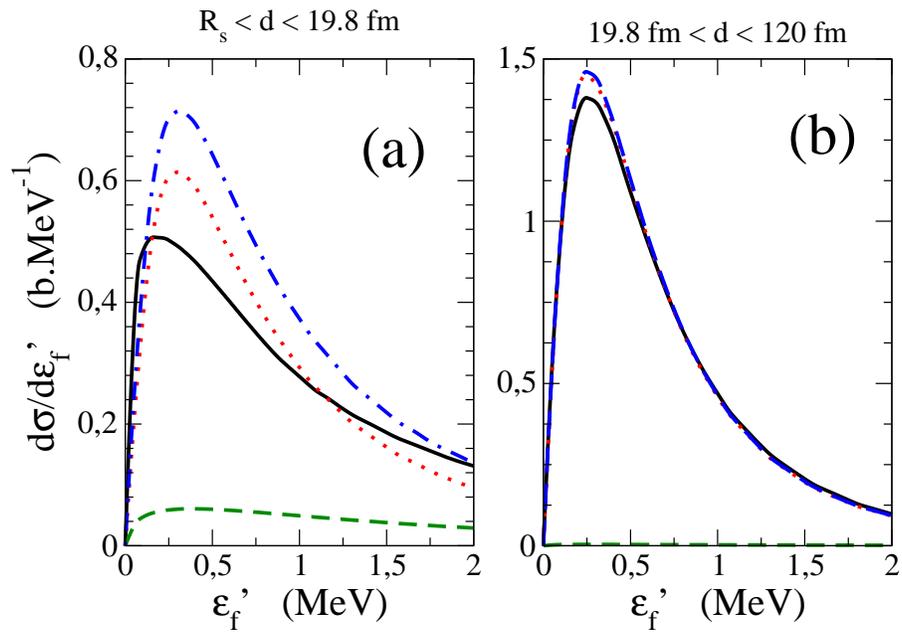}
\caption{Same as Fig.~(5) with  integration over impact parameters smaller and larger than
the critical value (cf. Table 1). The solid line is the $g_{\rm CN}^{\rm BBM\, 1}$ result,
  dotted is Coulomb  first order perturbation theory, dashed line is nuclear
  sudden to all orders and dotted-dash line is the previous result from
  \cite{mar02}.}
\label{fig6}
\end{figure}

The main results of our new calculations are shown in Fig.~(\ref{fig6}) (a) and
(b) by the solid thick line. The effective formula $g^{\rm BBM  \, 1}_{\rm CN}$ has been 
integrated in the impact parameter range smaller and larger than $d_{crit}=19.8$ fm
respectively.   These calculations indicate that for $R_s < d < 19.8$ fm the results
obtained with 
$g^{\rm BBM  \, 1}_{\rm CN}$, solid line, 
are smaller than those obtained with first order perturbation theory
$g^{\rm pert \, 1}_{\rm C}$, 
thus showing  higher order terms need to be considered. Since from the angular
distributions shown in the previous figures one can see that the effect of the higher
order terms in Coulomb breakup alone is rather small, we suggest that the strong depletion
shown by the peak of the energy distributions including Coulomb and nuclear breakup,
comes mainly from the destructive interference effect already discussed in \cite{mar02}. On
the other hand for
$d >19.8$ fm we find that higher order effects are negligible since using
the effective formula or perturbation theory  gives very little 
difference. We have checked that for 

\noindent $d >30$ fm perturbation theory agrees exactly with
the effective formula. Then we
can conclude that at large
$d$ perturbation theory is valid. It is then reasonable to think that experimental data for
neutron breakup could be analyzed by first order perturbation theory provided one could
extract the contribution from impact parameters somewhat larger than $d_{crit}$.  As we mentioned before,  it is important to notice that
the amplitude defined as $g^{\rm BBM \, 1}_{\rm C}$, valid at all core-target
impact parameters does  not give rise to any divergences in the final integral over
impact parameters. This is because the first order sudden term
$g^{\rm sudd\,1}_{\rm C }$ which contains the divergence is removed and substituted
by the first order time dependent perturbation theory  term $g^{\rm pert
\,1}_{\rm C}$ which does not diverge. The dashed line gives the nuclear contribution
alone and the dot-dashed line is the result  of  our previous method \cite{mar02},
recalculated using relativistic kinematics for consistency with all other results of this
paper. 
    
\begin{figure}[pht]
\center
\includegraphics[scale=0.5]{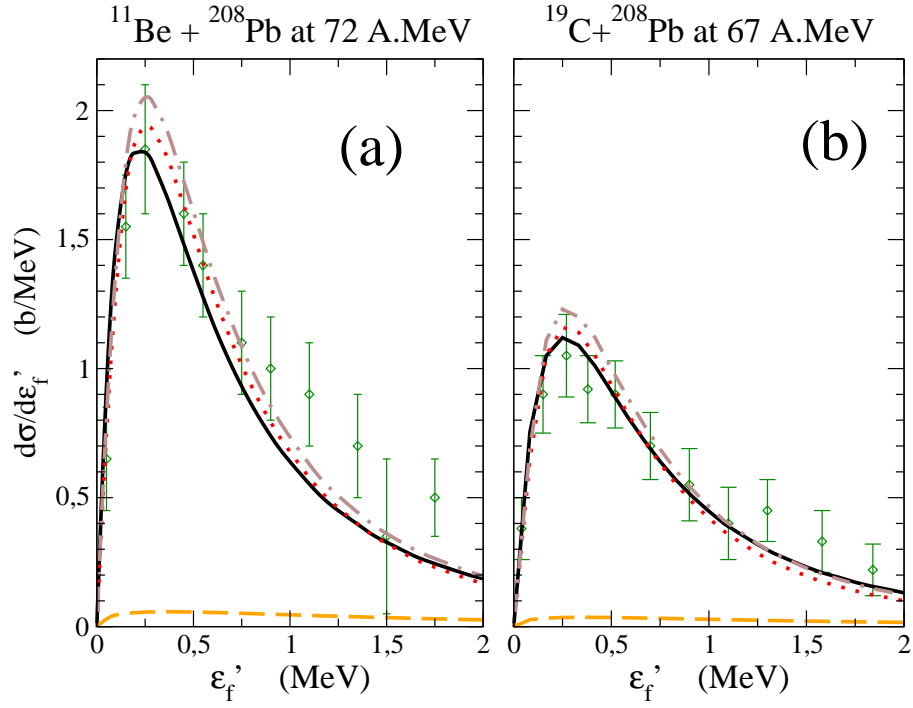}
\caption{Comparison between calculations and experimental results for the
  reactions: (a) $^{11}$Be+$^{208}$Pb at 72 A.MeV  and
(b) $^{19}$C+$^{208}$Pb
  at 67 A.MeV (right). The solid line is the $g^{\rm BBM  \, 1}_{\rm CN}$ result, dotted
is Coulomb  first
  order perturbation theory, dashed line is nuclear sudden to all orders  and
  dot-dashed line is previous results from \cite{mar02}. The spectroscopic
  factor  used  are $C^2S=0.77$ for $^{11}$Be and  $C^2S=0.65$ for $^{19}$C. Data are from
\cite{nak94,nak99}.}
\label{fig7}
\end{figure}

Finally the  results of the full impact parameter integration
are shown in Fig.~(\ref{fig7}) which gives the neutron
final energy spectrum with respect to the core  for breakup of
$^{11}$Be and $^{19}$C on $^{208}$Pb at 72 A.MeV and 67 A.MeV
respectively. Experimental data are from \cite{nak94,nak99}. Notations are as in Fig.~(6).
In the case of $^{11}$Be the theoretical calculations have been multiplied by
the known spectroscopic factor $C^2S=0.77$, while for $^{19}$C we have used
$C^2S=0.65$  and a neutron separation energy for
the 2s state of 0.5 MeV. As expected, and already shown by other authors the
effects of higher order terms are to reduce the peak cross section
\cite{typ01a}. Analysis of the type presented in this section have been used
to extract spectroscopic factors.

\section{Conclusions}

In this work we have presented an approximation scheme which allows
calculating Coulomb and nuclear breakup and their coupling to all orders.
It appears that higher order terms can give some effect only for heavy targets,
at low core-target impact parameters and small neutron angles, and their
effect is still noticeable after integrating over impact parameters. We have
shown that  higher order effects are accurately calculated by the sudden
approximation.  The neutron angular distribution observable is well
reproduced by the time dependent perturbation theory of Coulomb breakup.
Once again we would like to stress the usefulness of measuring neutron
angular distributions following breakup, as one of the best observables to
clarify reaction mechanisms and to test theoretical models. 

The neutron-core relative energy spectrum after breakup shows a depletion of
the peak value when higher order effects are included. It appears that this is mainly due
to the destructive interference between nuclear and Coulomb breakup.  Our conclusions are
in agreement with recent numerical solutions of the breakup problem
\cite{typ01a,typ01b,cha02} and with new experimental data \cite{nak3}. In particular as
higher order effects in the Coulomb  potential are important only for small impact
parameters ($d<d_{\rm crit}$),  in order to extract spectroscopic factors without any
ambiguities, it appears very well suited to compare Coulomb first order perturbation theory
with data containing contributions from
$d>d_{\rm crit}$ only \cite{nak3}.

One neutron halo breakup is a simple reaction mechanism for which our approximation scheme
allows to treat higher order terms in the Coulomb and the nuclear fields at the same time,
including their
 couplings. This approximation scheme leads to simple expressions which
can be generalized  to proton breakup, where one needs to
include the particle-target Coulomb potential and its quadrupole component.
\vskip .5cm
 {\bf Aknowledgments} 

We wish to thank Stefan Typel for discussions and comments  and for communicating his
results previuos to publication.

\appendix
\section{Eikonal perturbation theory for Coulomb 
breakup}

We give here some explicit formulae for the eikonal perturbation theory in the
case of an electromagnetic excitation. 
The Coulomb field from a target nucleus can act both on the core and on the
halo nucleus. Here we are only interested in the part that acts on their
relative position and cause the breakup. For this reason, we subtract the
part that acts on the position of the center of mass and we obtain
\cite{esb97} :
\be
V_{\rm eff} = Z_T
e^2\left(\frac{Z_h}{|\R+\beta_2\r|}+\frac{Z_C}{|\R-\beta_1\r|}
-\frac{Z_C+Z_h}{|\R|} \right)
\label{eq4}
\ee
where charges and masses are: core ($A_C$,$Z_C$), halo ($A_h$,$Z_h$), target
($A_T$,$Z_T$). We used also two ratios : $\beta_1=A_h/A_P$ and
$\beta_2=A_c/A_P$, with $A_P=A_C+A_h$. We develop this interaction in series
and we take the case of a zero charge halo thus getting the dipole field 
\be
V_{\rm eff}(\r,t) &=& f(t)x+g(t)z\label{eq5}
\ee
where 
\be
f(t)=\frac{Z_{\rm eff} d}{\big(d^2+(vt)^2\big)^{3/2}} 
\hspace{1cm} {\rm and} \hspace{1cm}
g(t)=\frac{Z_{\rm eff} vt}{\big(d^2+(vt)^2\big)^{3/2}}. \nonumber
\ee
and $Z_{\rm eff}=\beta_1Z_PZ_Te^2$.
According to the eikonal formalism, the phase shift is expressed in the
simple form :
\be
\chi_{\rm eff }(\r,t) =\int_t^\infty V_{\rm  eff}(\r,t^{\prime})
dt^{\prime} &=& F(t) x+G(t) z
\label{eq3}
\ee
with
\be
F(t) &=&{1\over \hbar} \int_t^\infty f(t^{\prime}) dt^{\prime} 
={Q\over 2}\left(1-\frac{s}{\sqrt{1+s^2}}\right)
\nonumber \\
G(t) &=& {1\over \hbar}\int_t^\infty g(t^{\prime}) dt^{\prime}
={Q\over 2}\frac{1}{\sqrt{1+s^2}}
\label{eq4a}
\ee
where $s=vt/d$ and $Q=2 Z_{\rm  eff}/(\hbar v d)$.

Introducing Eq.~(\ref{eq3}) into Eq.~(\ref{amp2}) leads to
\be
g^{\rm pert}_{\rm C}(\k,\d;lm) 
%&=& -\inti dt \, \e^{i{\omega}t} \frac{d}{dt}
%\!\!\inti d^3\r\,\e^{-i\k\cdot\r-i\chi_{\rm eff}(\d,t)}\,
%\phi_{lm}(\r) \nonumber \\ 
&=& -\inti dt\,\e^{i{\omega}t} \frac{d}{dt}
\tilde{\phi}_{lm}(\k_t)
%\label{eq5}
\ee
where $\k_t=\big(k_x+F(t),k_y,k_z+G(t)\big)$.  Using Eq.~(\ref{f}) the breakup
amplitude from an initial s-state becomes
\be
g^{\rm pert}_{\rm C}(\k,\d;00) 
%&=& -\int dt\e^{i\omega t} \frac{d}{dt}\int d^3\r
%\e^{-i\k_t\cdot\r} \phi_{00}(\r) \label{eqq5} \\
=4 C_i \sqrt{\pi} \int dt\,\e^{i\omega t} \frac{f(t)[k_x+F(t)]+g(t)[k_z+G(t)]}
{\left([k_x+F(t)]^2+{k_y}^2+[k_z+G(t)]^2+\gamma_0^2\right)^2}.
\label{eqq6}
\ee

We obtain the eikonal perturbation theory  from an expansion in powers of the
phase shift. Due to the simple form of the phase shift $\chi_{\rm  eff}=Qx$, we
expand Eq.~(\ref{eqq6}) with respect to the small quantities $F(t)k_x/K^2$ and
$G(t)k_z/K^2$ which are proportional to the parameter $\xb=Q/K$ used in
Eq.~(\ref{g}), 
and neglect terms of order $F(t)^2$ and $G(t)^2$. Thus the first order term
reads : 
\be
g^{\rm pert \, 1}_{\rm C}(\k,\d;00)
%&=& \frac{4 C_i \sqrt{\pi}} {K^2} \left( k_x \int dt \,\e^{i\omega t} f(t)  +
%  k_z \int dt\, \e^{i\omega t} g(t)\right)  \nonumber \\
&=&-\sqrt{\pi}  C_i Q \frac{\wb} {K^4}
\Big(2k_xK_1(\wb)+ 2ik_zK_0(\wb)\Big)\label{eq40}
\ee
where  one sees that
the adiabaticity parameter $\wb$ appears naturally. The second order amplitude is
\be
g^{\rm pert \, 2}_{\rm  C}(\k,\d;00) 
%&=&\frac{4 C_i \sqrt{\pi}} {K^6}\Big\{\rm K^2 \int dt \,\e^{i\omega t} \Big(
%f(t)F(t) + g(t)G(t) \Big) \nonumber \\ 
%&&-4 \int dt \,\e^{i\omega t} \Big( k_xf(t)+k_zg(t)\Big)\Big( k_x F(t)+k_zG(t)
%\Big) \Big\}\nonumber\\
&=&- {\sqrt{\pi}\over 2} C_i  Q^2\frac{\wb}{K^6}
\Big( \big( K^2-4{k_x}^2 \big ) K_1(\wb) -4ik_x k_zK_0(\wb) \nonumber \\
&&
-i\pi({k_z}^2+2k_x k_z-{k_x}^2)\e^{-\wb} \Big)\label{eq41}
\ee
The sudden approximation of these amplitudes are simply deduced as the limit
$\wb\rightarrow 0$.

In order to make a link with previous work 
\cite{typ01a} by other authors, we calculate explicitly the
probability momentum distribution in the sudden approximation as:
\be
\frac{d P^s(\d)}{d k}=\frac{1}{8\pi^3}\int d\Omega^{\prime} {k}^2
|g^{\rm sudd}_{\rm C}(\k,\d)|^2
\ee
Then, we find for the first order and second order terms in the dipole field
\be 
\frac{d P_1^{s}(\d)}{d k} &=& \frac{1}{8\pi^3}\int d\Omega^{\prime}{k}^2
|g^{\rm sudd\,1}(\k,\d)|^2 =\frac{16y^2}{3\pi\gamma_0}\frac{x^4}{(1+x^2)^4}
\label{eq42}\\
\frac{d P_{22}^{s}(\d)}{d k} &=& \frac{1}{8\pi^3}\int d\Omega^{\prime}{k}^2
|g^{\rm sudd\,2}|^2  =\frac{4y^4}{15\pi\gamma_0}
\frac{x^2(15-10x^2+23x^4)}{(1+x^2)^6}\label{eq43} \\
 \frac{d P_{13}^{s}(\d)}{d k} &=& \frac{1}{8\pi^3}\int d\Omega^{\prime}{k}^2
g^{\rm sudd\,1}g^{\rm sudd\,3} =-\frac{32y^4}{15\pi\gamma_0}
\frac{x^4(5-x^2)}{(1+x^2)^6}\label{eq44}\\
% \frac{d P_{12}^{s}(\d)}{d k} &=& \frac{1}{8\pi^3}\int d\Omega^{\prime}{k}^2
%g^{\rm sudd\,1}g^{\rm sudd\,2} = 0\\
\frac{d P_{2}^{s}(\d)}{d k}& = &\frac {d P_{22}^{s}(\d)}{d k} 
+2\frac{d P_{13}^{s}(\d)}{d k}
=\frac{4y^4}{15\pi\gamma_0} \frac{x^2(15-90x^2+39x^4)}{(1+x^2)^6}
\ee
with $x=k/\gamma_0$, $y=Q^{(D)}/\gamma_0$ and we used the 
asymptotic normalization constant $C_i^2= 2\gamma_0$ of authors \cite{typ01a}.

\begin{figure}[pht]
\center
\includegraphics[scale=.5, angle=90]{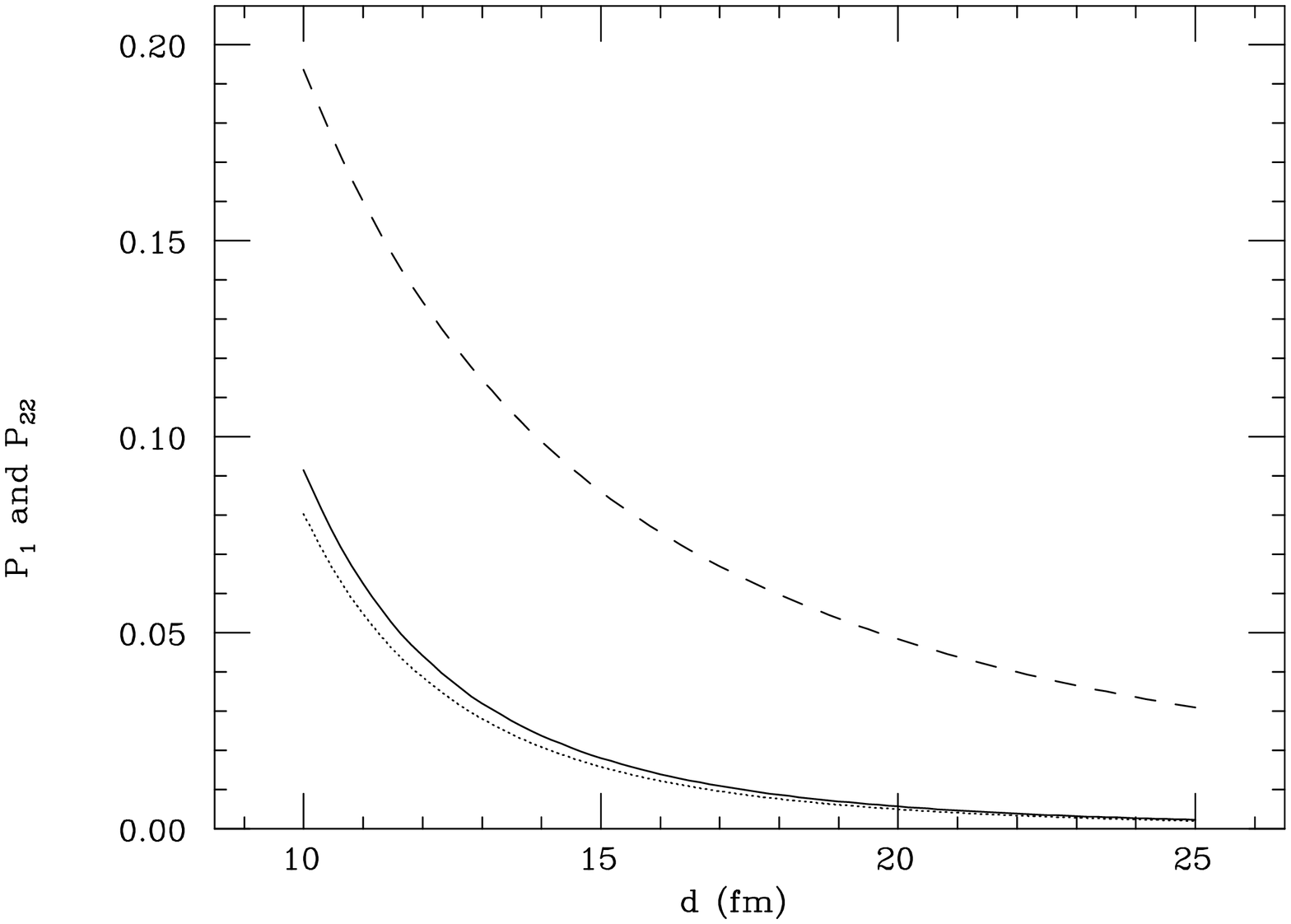}
\caption{Comparison between the sudden first order Coulomb breakup probability
 Eq.~(\ref{eq42}) after momentum integration, as a function of 
the  core-target impact parameter (dashed line) and the second order
dipole-dipole term 
calculated according to Eqs.~(\ref{amp1}) and (\ref{eq43}) (solid line) and 
with the final plane wave function substituted by Eq.~(\ref{eq45}) (dotted
line). }
\label{fig8}
\end{figure}

Our first order result Eq.~(\ref{eq42}) is identical with the equivalent term
Eq.~(2.6) of  
ref.\cite{typ01a}. The second order dipole-dipole term  of this work
Eq.~(\ref{eq43}) is 
different from Eq.~(2.8) of 
\cite{typ01a}. The difference is due to the fact that in \cite{typ01a}  the exact
neutron-core scattering wave function 
\be 
\phi_f({\bf k})={\e^{i{\bf k \cdot r}}}
-{1\over{\gamma_0-ik}}{e^{ikr}\over 
r}.
\label{eq45}
\ee
was used, while in this paper   the plane wave approximation has been applied,
consistently with our hypothesis that the neutron-core final state interaction can be
neglected (cf. Sec. 2 of
\cite{mar02}). We have checked that using Eq.~(\ref{eq45}) we would get the same result as
in
\cite{typ01a} but 
also that the difference with Eq.~(\ref{eq43}) is negligible. To show this
point we give in 
Fig.~(\ref{fig8}) the
  sudden first order Coulomb breakup probability Eq.(\ref{eq42}) after momentum integration
as a function of the  core-target impact parameter $d$ (solid line) and the second order
dipole-dipole term Eq.~(\ref{eq43})
calculated according to Eq.~(\ref{amp1}) (dashed line) and  with the final
plane wave 
function substituted by Eq.~(\ref{eq45}) (short dashed line). It is clear that
second order 
terms are rather small compared to the first order term, but also that the use
of final plane 
waves is justified.

\section{Fourier Transforms for the Coulomb potential}

\be
\int dt \,\e^{i\omega  t} f(t) &=&  Q \wb K_1(\wb)\\
\int dt \,\e^{i\omega  t} g(t) &=&  i Q \wb K_0(\wb)
\ee

\be
\int dt \,\e^{i\omega  t} f(t) F(t)&=& {Q^2\over 4}
\left(2\wb K_1(\wb)-i\frac{\pi}{2}\wb\e^{-\wb}\right) \\
\int dt \,\e^{i\omega  t} g(t) G(t)&=& {Q^2\over 4} i\frac{\pi}{2}\wb\e^{-\wb}\\
\int dt \,\e^{i\omega  t} f(t) G(t)&=& {Q^2\over 4} \frac{\pi}{2}(1+\wb)\e^{-\wb}\\
\int dt \,\e^{i\omega  t} g(t) F(t)&=& {Q^2\over 4}
\left(2i\wb K_0(\wb)+\frac{\pi}{2}(-1+\wb)\e^{-\wb}\right)
\ee

\begin{thebibliography}{99}

\bibitem{typ01a} S. Typel and G. Baur, \pr{C 64} (2001) 024601.

\bibitem{typ01b} S. Typel and R. Shyam, \pr{C 64} (2001) 024605.

\bibitem{tf} H. Esbensen, G. Bertsch and C.A. Bertulani, \np{A581} (1995) 107.

\bibitem{kys} T. Kido, K. Yabana and Y. Suzuki, Phys. Rev. C {\bf 53 } (1996) 2296.

\bibitem{mb} V. S. Melezhik and D. Baye, Phys. Rev. C {\bf 59}  (1999) 3232.

\bibitem{fal01} M. Fallot, J. A. Scarpaci, D. Lacroix, Ph. Chomaz and
  J. Margueron, \np{A700} (2002) 70.

\bibitem{esb01}  H. Esbensen, and G. F. Bertsch,  Phys. Rev. C {\bf 64} (2001) 014608;
\np{A706} (2002) 383.

\bibitem{cha02} R. Chatterjee and R. Shyam,   Phys. Rev. C {\bf 66} (2002) 061601.

\bibitem{nt} F. M. Nunes and I. J. Thompson, Phys. Rev. C {\bf 59}, (1999) 2652  and
references therein.


\bibitem{bon01} A. Bonaccorso and G. F. Bertsch, \pr{C 63} (2001) 044604.

\bibitem{mar02} J.Margueron, A.Bonaccorso and D.M. Brink, \np{A703} (2002) 105
  and references therein.

\bibitem{bon99} A. Bonaccorso, \pr{C 60} (1999) 054604 and references therein.



\bibitem{ald75} K. Alder and A. Winther, electromagnetic excitation 
(North-Holland, Amsterdam, 1975).


\bibitem{typ94a} S. Typel and G. Baur, \pr{C 49} (1994) 379. 

\bibitem{typ94b} S. Typel and G. Baur, \pr{C 50} (1994) 2104. 

\bibitem{typ94c} S. Typel and G. Baur, \np{A573} (1994) 486.

\bibitem{typ95} S. Typel and G. Baur, \pl{B 356} (1995) 186.

\bibitem{tok} Y.Tokimoto et al., \pr{C 63} (2001) 35801.

\bibitem{esb97} H.Esbensen, International school of heavy ion physics, Erice
  11-20 May 1997, Ed R.A.Broglia and P.G.Hansen, World Scientific.



\bibitem{bon98a} A. Bonaccorso and D. M. Brink, \pr{C 57} (1998) R22.

\bibitem{bon98b} A. Bonaccorso and  D. M. Brink, \pr{C 58} (1998) 2864. 

\bibitem{win79} A. Winther and K. Adler, \np{A319} (1979) 518. 

\bibitem{ann94} R. Anne et al., \np{A575} (1994) 125. 

\bibitem{nak94} T. Nakamura, \pl{B 331} (1994)  296. 

\bibitem{nak99} T. Nakamura et al., \prl{83} (1999) 1112.

\bibitem{nak3} T. Nakamura,  in Proc. 
of the 4th Italy-Japan Symposium on 
Perspectives in Heavy Ion Physics, edited by
K. Yoshida, S. Kubono, I. Tanihata, C. Signorini, 
(World Scientific, 2003), p. 25, and to be published. 

\end{thebibliography}
\end{document}